\documentclass[authoryear,preprint,review,12pt]{elsarticle}
\usepackage{multirow}
\usepackage{graphicx}
\DeclareGraphicsExtensions{.pdf,.jpeg,.png}
\usepackage{epstopdf}
\usepackage{subfigure}
\usepackage{booktabs}
\usepackage{mathrsfs}
\usepackage{rotating}
\usepackage{lscape}
\usepackage{threeparttable}
\usepackage{natbib}
\usepackage{amsmath}
\usepackage{amssymb}
\usepackage[colorlinks,linkcolor=red,anchorcolor=blue,citecolor=green]{hyperref}

\newcommand       \g          {\,{\rm g}}
\newcommand       \cm          {\,{\rm cm}}

\newcommand       \mum           {\,{\rm \mu m}}
\newcommand       \Rv           {{R_{\rm V}}}
\newcommand       \Av           {{A_{\rm V}}}

\newcommand       \magni        {\,{\rm mag}}

\newcommand       \nH           {n_{\rm H}}
\newcommand       \NH           {N_{\rm H}}

\journal{Journal of \LaTeX\ Templates}

\begin{document}

\begin{frontmatter}

\title{Dust Extinction towards the Type Ia Supernova 2012cu in NGC 4772}

\author[1]{WEIJIA GAO}
\author[1]{RUINING ZHAO}
\author[1]{JIAN GAO\corref{cor1}}
\ead{jiangao@bnu.edu.cn}
\author[1]{BIWEI JIANG}
\author[1]{JUN LI}
\address[1]{Department of astronomy, Beijing Normal University, Beijing 100875, China}
\cortext[cor1]{Corresponding author}

\begin{abstract}
Using photometric and spectroscopic data of Supernova (SN)
2012cu, a fairly reddened type Ia supernova,
we derived its color excess curves and probed the dust
extinction in its host galaxy, NGC 4772.
In order to derive the extinction as a function of wavelength
(i.e., $A_\lambda$), we model the color excess curves of
SN 2012cu in terms of dust models consisting of silicate
and carbonaceous (graphite or amorphous carbon) dust.
The modeled extinction law towards SN 2012cu extends
flatly to the far-ultraviolet (UV) bands,
which is much flatter than those of the Milky Way
and Magellanic Clouds, and the 2175{\AA} feature is very weak or absent.
The flatness of the modeled extinction
curve in the UV bands suggests a ``grey'' extinction
law of the active galactic nucleus in the vicinity of
the SN 2012cu-Earth line of sight.
Our results indicate that the sizes of the dust in the
ISM towards SN 2012cu in NGC 4772 are larger than
those of the Milky Way and the Large Magellanic Cloud,
and much larger than that of the Small Magellanic Cloud.
The best fitting gives an observed visual extinction
towards SN 2012cu of $\Av \approx 2.6\magni$, a reddening of
$E(B-V)\approx1.0\magni$, with a total-to-selective extinction
ratio $\Rv$\,$\approx$\,2.7, consistent with previous results.
\end{abstract}

\begin{keyword}
(ISM:) dust, extinction {\textendash} galaxies: ISM {\textendash} galaxies: individual (NGC 4772) {\textendash} supernovae: individual (SN 2012cu)
\end{keyword}

\end{frontmatter}

\section{Introduction}\label{sec:intro}
Attributed to their high luminosity with small dispersion at
the maximum of light curves, type Ia supernovae (SNe Ia) are
one of the most powerful ``standard candles'' for measuring
cosmological distances (\citealt{Riess98,Riess16,Perlmutter}),
leading to the discovery of accelerating universe and hence
the presence of mysterious dark energy.
Incomprehensive knowledge on the extinction towards
SNe Ia will, however, result in a systematic uncertainty in
the intrinsic luminosity and distances to SNe Ia, which
leads to one of the major issues in SNe Ia cosmology as the
correction for the line-of-sight extinction.
In addition, due to the undetectable extragalactic
environment, SNe Ia are often the only approach to study the
extinction of extragalactic dust.

The wavelength dependence of the interstellar extinction
[known as the ``interstellar extinction law (or curve)'']
is defined as $A_{\lambda}$ at wavelength $\lambda$.
Since $A_{\lambda}$ is hard to obtain directly,
astronomers often measure the color excesses $E(\lambda-V)\equiv A_\lambda - \Av $,
where $\Av$ is the extinction in the visual band.
The color excesses of SNe Ia are often determined by comparing
spectrophotometry of two sources with the same spectral shape, one of which
has minor foreground reddening (i.e., \citealt{A14,A15}).
The total-to-selective extinction ratio
$\Rv\equiv \Av/E(B-V)$ provides an adequate
description of the extinction laws of the Milky Way (MW) dust
(e.g., \citealt[hereafter \hyperlink{CCM89}{CCM89}]{Cardelli};
\citealt[hereafter \hyperlink{F99}{F99}]{F99}), in which
$\Rv$ ranges from 2.2 to 5.5 with the average value
of 3.1 (e.g., \citealt{FM07}).
However, \hyperlink{CCM89}{CCM89} and \hyperlink{F99}{F99}
are not necessarily valid for the extinction laws for external
galaxies, even for the Large and Small
Magellanic Clouds (LMC, SMC; \citealt{Gordon}).

There is increasing evidence that extinction curves
towards SNe Ia systematically favor a steeper law
($\Rv < 3$, see, for instance, \citealt{Nobili,Folatelli,Cikota})
compared to the Galactic average value
($\Rv = 3.1$, see \hyperlink{CCM89}{CCM89} or \hyperlink{F99}{F99}).
This suggests the properties of extragalactic dust may be
incompatible with the Milky Way dust.
Additionally, scattering by circumstellar dust also tends to
reduce $\Rv$ in the optical band
(\citealt{Wang05,Patat07,Goobar08}).
This discrepancy leads to another one of the major issues in SNe\,Ia
cosmology, which is to understand whether the systematically
low $\Rv$ values towards SNe\,Ia are caused by 1) systematic
differences from the optical properties of extragalactic dust
grains, or 2) modifications by the circumstellar matter scattering.

In our previous work, \citet{Gao} found that the peculiar
extinction law in the sightline of SN 2014J, one SN Ia with
$\Rv < 2$ (e.g., $\Rv = 1.6 \pm 0.2$, \citealt{Foley}),
cannot be explained by \hyperlink{CCM89}{CCM89}. For comparison,
the dust models of a mixture of silicate and graphite/amorphous
carbon dust grains provide an excellent fitting to the observed color excess
curves towards SN 2014J. The reddening curve fitting around the
peak luminosity of SN 2014J gives $\Rv \sim 1.7$ towards the SN 2014J line
of sight, which is also generally consistent with many other studies
(\citealt{A14,Foley,Goobar,Brown,Yang}).

SN 2012cu was first discovered on June 11.2 UT, 2012 by \citet{Itagaki}
and later classified as a SN Ia on June 15 UT, 2012 by \citet{Marion}.
It locates at ($\alpha_{2000},\delta_{2000}$) = (12:53:29.35, +02:09:39.0),
$3^{''}.1$ east and $27^{''}.1$ south of the nucleus of
the host galaxy NGC 4772 at a distance of $15.6 \pm 1.0$ Mpc\footnote{The
measured distances to NGC 4772 show the discrepancy among previous studies.
\citet{A15} adopted a distance of $41 \pm 9$ Mpc derived from the
distance modulus of $33.06 \pm 0.36 \magni$ for NGC 4772 (see \citealt{Tully08} and Table 1 therein),
while \citet{H17} obtained the distance modulus of $31.11 \pm 0.15 \magni$ corresponding to a
distance of $16.6 \pm 1.1$ Mpc. \citet{H17} thought that the former measured with Tully-Fisher relation
would be too large due to the bulk of the gas in NGC 4772.}, which can be found in the
Extragalactic Distance Database (EDD, \citealt{Tully})\footnote{http://edd.ifa.hawaii.edu/}.
The host galaxy is a Sa-type spiral galaxy according to its morphology
(\citealt{Haynes}), and also classified as a low-luminosity ``dwarf''
Seyfert nuclei (or low-ionization nuclear emission-line region, LINER)
according to its spectral lines (\citealt{Ho}).
SN 2012cu is regarded as one of the reddest SNe Ia with $E(B-V) \sim 1 \magni$.
In the previous studies, \citet[hereafter \hyperlink{A15}{A15}]{A15}
derived $E(B-V) = 0.99 \pm 0.03 \magni$ and $\Rv = 2.8 \pm 0.1$
of SN 2012cu by fitting its UV-to-near-infrared (NIR) photometric colors.
\citet[hereafter \hyperlink{H17}{H17}]{H17} found their best-fit
$E(B-V) = 1.00 \pm 0.03 \magni$ and $\Rv = 2.95 \pm 0.08$
by dereddening its optical spectra at different epochs.
Both \hyperlink{A15}{A15} and \hyperlink{H17}{H17} obtained the $\Rv$
values of SN 2012cu by using the $\Rv$-based formula of \hyperlink{F99}{F99}.

In this work, using the photometric data from \hyperlink{A15}{A15}
and spectroscopic data from \hyperlink{H17}{H17} (see \S \ref{sec:obda}),
we measure the color excesses $E(\lambda-V)$ of SN 2012cu.
Without making a priori assumption of any template extinction law,
we derive $\Rv$ and $A_\lambda$ of SN 2012cu from the far-UV to NIR
bands, in terms of mixed dust models of silicate and graphite or
amorphous carbon (see \S \ref{sec:model}).
The results are presented in \S \ref{sec:result},
discussed in \S \ref{sec:diss}, and summarized in \S \ref{sec:conc}.

\section{Observational Data}\label{sec:obda}
We utilize the photometric data of SN 2012cu published by
\hyperlink{A15}{A15}\footnote{All data and figures presented by
\hyperlink{A15}{A15} are available at http://snova.fysik.su.se/dust/.}.
Their observations were performed with {\it Hubble Space Telescope}
({\it HST})/Wide-Field Camera 3 (WFC3) through passbands of $F225W$,
$F275W$ and $F336W$; and Nordic Optical Telescope (NOT, \citealt{Djupvik})
/Andalucia Faint Object Spectrograph and Camera (AL-FOSC) in $U$,
$B$, $V$, $R$ and $i$, and NOTCam in the $J$, $H$ and $Ks$ bands.
The wavelength of the observed spectral energy distribution (SED)
ranges from 2346 to 21295{\AA} (i.e., from the UV to NIR bands), and
uncertainties range from 0.08 to 0.3 mag (e.g., $\sim 0.08 \magni$ for
$R$ filter, $\sim 0.2 \magni$ for $Ks$ band and $\sim 0.3 \magni$
for $F225W$ and $F275W$ passbands, \hyperlink{A15}{A15}).
In total, 31 measurements are collected at eight epochs\footnote{The
epoch is the day relative to the $B-$band maximum light
of the observation.} (see \hyperlink{A15}{A15} and Figure 6 therein).
The colors of SN 2012cu are obtained by making the redshift corrections
(K-corrections, e.g., \citealt{Hamuy}) and magnitude corrections (S-corrections,
e.g., \citealt{Stritzinger}), and subtracting the MW foreground extinctions
($A^{\rm MW}$) with the following equation (\hyperlink{A15}{A15})
\begin{equation}
X - V = (m_X - A_X^{\rm MW} - K_X) - (m_V - A_V^{\rm MW} - K_V),
\label{eq1}
\end{equation}
where $X$ ($V$) is the rest-frame magnitude in the filter X (V);
$m_X$ and $m_V$ are the observed magnitudes; $K_X$ is the combined K
and S corrections; $A^{\rm MW}$ and $K_X$ are given by \hyperlink{A15}{A15},
who adopted the values of \citet{Schlafly}.
Then the color excesses of SN 2012cu can be derived by using the following
equation
\begin{equation}
E(X - V) = (X - V) - (X - V)_0,
\label{eq2}
\end{equation}
where $(X - V)_{0}$ is the intrinsic color of the unreddened objects listed
by \hyperlink{A15}{A15}, who combined the spectroscopic data of SN 2011fe
and created a series of daily sampled unreddened SED templates to
derive the intrinsic colors of SNe Ia. It will be described later in this
section.

The spectroscopic data of SN 2012cu are taken from \hyperlink{H17}{H17}\footnote
{The spectra of SN 2012cu are available at
http://snfactory.lbl.gov/TBD.} (see Figure 13 therein),
which span 17 epochs from -6.8 days to 46.2 days.
In this work, we use the spectra which span 15 epochs from -6.8 days to
31.2 days\footnote{The epochs at 41.2 and 46.2 days are excluded because
the SED templates of SN 2011fe used in this work only extend to 40 days
past $B-$band maximum (see \hyperlink{A15}{A15} and APPENDIX B therein).}.
The spectra of SN 2012cu cover a wavelength range from 3300 to 9200{\AA},
and have already been flux calibrated (\citealt{Buton}), host-galaxy
subtracted and corrected for the MW foreground extinctions
(\citealt{Schlafly}).

In order to determine the intrinsic colors of SN 2012cu,
we use the SED templates of SN 2011fe privately provided by
Amanullah R. (\hyperlink{A15}{A15}).
SN 2011fe, one of the best studied SNe Ia,
is the rare one with high-quality UV time series spectra
obtained with {\it HST} (\citealt{Foley,Mazzali}).
Furthermore, the extremely low host reddening
(i.e., $E(B-V)_{\rm host} = 0.026 \pm 0.036 \magni$, \citealt{Nugent,Pereira,Zhang})
and the absence of complex absorption profiles (\citealt{Patat}) towards
SN 2011fe suggest negligible extinction from its host galaxy.
Hence, with the wide wavelength and comprehensive phase coverage
dataset, the nearby, reddening-free SN 2011fe provides an
excellent calibration for extinction comparison.

\begin{figure}[htbp]
 \center{\includegraphics[width=10cm]  {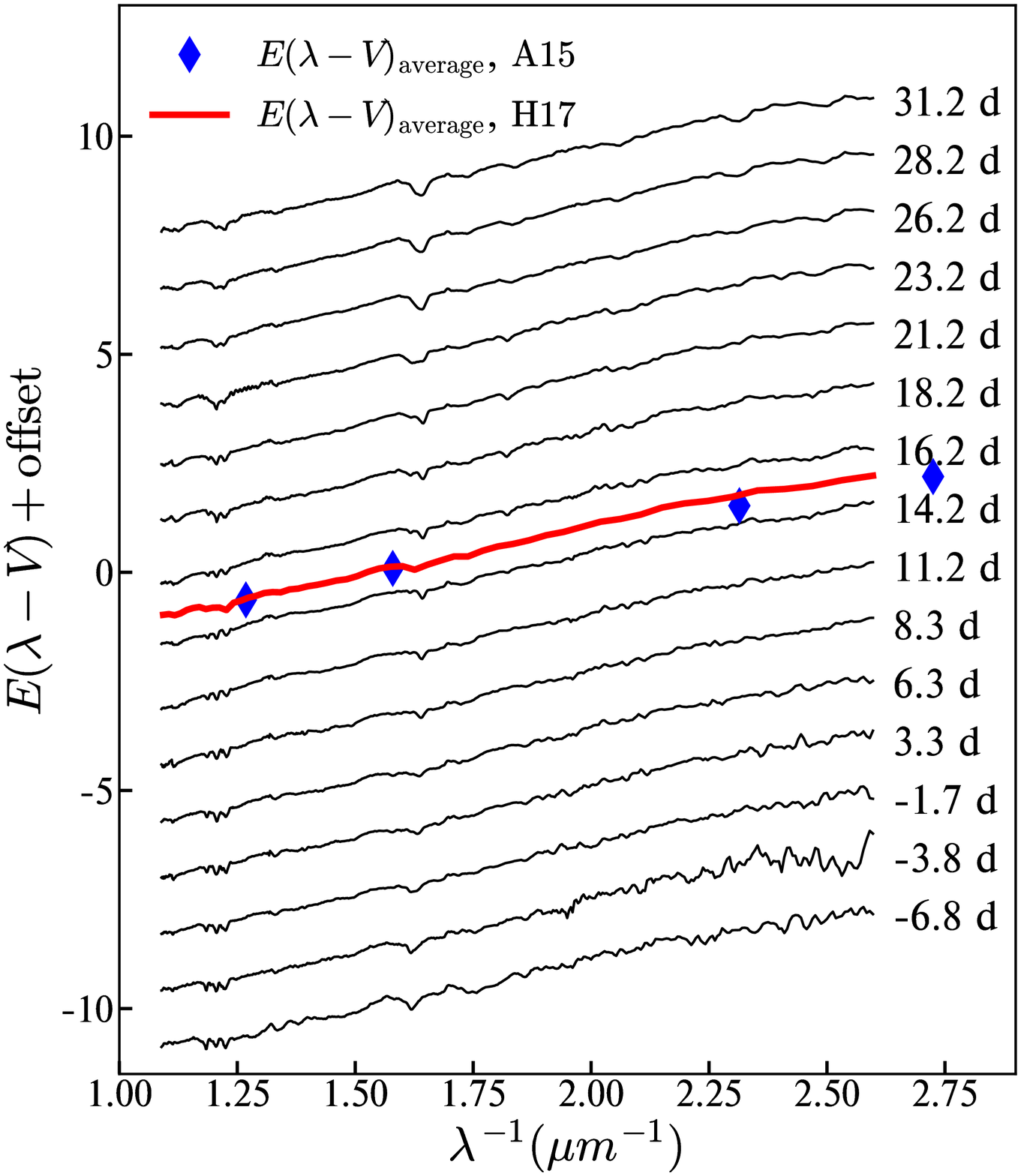}}
 \caption{\label{2} The derived color excess curves from the observed spectra of
 SN 2012cu.
 Black solid lines are the color excess curves of 15 epochs obtained from spectra.
 For comparison, the phase-averaged color excess curve is plotted in red solid
 line and the phase-averaged color excesses from photometric data at corresponding
 wavelength are shown in blue diamonds.
}
\end{figure}

Assuming that SN 2012cu has the same SED as the lightly reddened
SN 2011fe, the color excess curves of SN 2012cu can be directly derived
by comparing the spectra of SN 2012cu with the SED templates of SN 2011fe
at similar epochs (\hyperlink{A15}{A15}), as
\begin{equation}
E(\lambda - V)_{p} = -2.5[\log(\frac{f({\lambda;p})}{S_{0}(\lambda;p)}) - \log(\frac{f(V;p)}{S_0(V;p)})],
\label{eq3}
\end{equation}
where at wavelength $\lambda$ for phase \emph{p}, $f(\lambda;p)$ and $S_{0}(\lambda;p)$ are
flux densities from the spectra of SN 2012cu and the SED templates of
SN 2011fe, respectively.
Figure \ref{2} shows the color excess curves derived from the spectroscopic data.
These color excess curves are smoothed by a simple moving average
method. The substantially fluctuant parts of curves beyond wavenumber
$\geq 2.6{\mu m}^{-1}$ are also excluded.
%
%
%
\section{Dust Models}\label{sec:model}
We adopt a two-component grain model consisting of astronomical silicate
and graphite (GRA, \citealt{Draine}) or amorphous carbon (AMC, \citealt{Rouleau}).
We assume that both the silicate and carbonaceous grains have the same
size distribution, i.e., a power-law function with an exponential cutoff
(\citealt{Kim}; \citealt{Wang}; \citealt{Gao}):
\begin{equation}
\frac{dn_i}{da} =  B_i\nH a^{-\alpha}\exp(-a/a_b),
\label{eq4}
\end{equation}
where $\alpha$ and $a_b$ are the power index and exponential cutoff radius,
respectively; $dn_i$ is the number density of dust species $i$ (silicate
or carbonaceous component) with radii between $[a,a+da]$;
$n_{\rm H}$ is the number density in unit of hydrogen nucleon H;
$B_i$ is the normalization constant.
A classical \hyperlink{MRN}{MRN} grain model (\citealt[hereafter \hyperlink{MRN}{MRN}]{MRN})
has a power-law size distribution, i.e., $dn(a)/da \propto a^{-3.5}$ (\citealt{Clayton}).
When $a_b \gg a$, the exponential cutoff will have $\exp(-a/a_b) \approx 1$,
which decays Equation (\ref{eq4}) to the \hyperlink{MRN}{MRN} size distribution.

Based on the equations in \citet{Wang} (see Section 3 therein), the factor $B_{i}$
can be derived from
\begin{equation}
B_{\rm sil} = 1/(\rho_{\rm sil} \int da(4\pi/3)a^3a^{-\alpha}\exp(-a/a_{b})), \\
\label{eq5}
\end{equation}
for silicate dust, and
\begin{equation}
B_{\rm car} = f_{\rm cs}/(\rho_{\rm car} \int da(4\pi/3)a^3a^{-\alpha}\exp(-a/a_{b})), \\
\label{eq6}
\end{equation}
for carbonaceous dust (i.e., graphite or amorphous carbon),
where $f_{\rm cs}$ is the mass ratio of carbonaceous dust to silicate dust,
i.e., $m_{\rm car}/m_{\rm sil}$. The mass densities of amorphous silicate
($\rho_{\rm sil}$) and graphite ($\rho_{\rm GRA}$) are
the same as those used in \citet[hereafter \hyperlink{WD01}{WD01}]{WD01},
i.e., $3.5\g\cm^{-3}$ and $2.24\g\cm^{-3}$, respectively.
The density of amorphous carbon ($\rho_{\rm AMC}$) is $1.8\g\cm^{-3}$ (\citealt{Draine2011}).

Detailed information on the abundance of dust grains in the host galaxy of
SN 2012cu remains uncertain. Therefore, we adopt three typical values of
$f_{\rm cs}=0, 0.3$, and 0.6. When $f_{\rm cs}=0$, it means that there are
no carbonaceous grains in the dust model.
By assuming that the interstellar abundances in NGC 4772 are similar to
those of the MW and adopting the abundance values of \citet{Asplund},
the other two values of $f_{\rm cs}$ are derived.
The value of $f_{\rm cs}$ will be 0.6 when the elements of Fe, Mg and
Si are all in solid phase and constrained in the silicate dust,
and the fraction of gas-phase carbon is 0\% (\citealt{Gao13,Gao}; \citealt{Wang}).
In addition, $f_{\rm cs}$ will be 0.3 when the fraction of gas-phase carbon is 50\%.

Then the modeled extinction $A^{}_{\lambda}$ at wavelength $\lambda$ is calculated by
%
\begin{equation}
A^{}_{\lambda} = 1.086{\NH}\sum\limits_{i} \int^{a_{\rm max}}_{a_{\rm min}} da \frac{1}{\nH} \frac{dn_i}{da} C_{{\rm ext},i}(a,\lambda),
\label{eq8}
\end{equation}
where $\NH$ and $\nH$ are the column density and volume density in unit of
hydrogen nucleon H, respectively. All dust grains are assumed to be
spherical with the radius of $0.005\mum = {a_{\rm min}} \leq a \leq {a_{\rm max}} = 5 \mum$.
The extinction cross section $C_{{\rm ext},i}(a,\lambda)$ ($\rm cm^{-2}$), for
the grain of species \emph{i} with size $a$ at wavelength $\lambda$, is calculated
with the Mie theory code derived from BHMIE (\citealt{Bohren}).
The optical constants of dust are taken from
\citet{Draine} for astronomical silicate and graphite,
and from \citet{Rouleau} for amorphous carbon.

The modeled extinction $A_{X,p}$ in X band for phase $p$ is subsequently
derived from the similar formula by \hyperlink{A15}{A15}:
%
\begin{equation}
A_{X,p} = - 2.5\log[\frac{\int T_{X}(\lambda)\times10^{-0.4\times A_{\lambda}} \times S_{0}(\lambda;p)\lambda d\lambda}{\int T_{X}(\lambda)\times S_{0}(\lambda;p)\lambda d\lambda}],
\label{eq9}
\end{equation}
%
where the extinction $A_{\lambda}$ is initially calculated by Equation (\ref{eq8});
$T_{X}(\lambda)$ is the filter transmission,
and $S_{0}(\lambda;p)$ is the unreddened flux density of the SED at wavelength
$\lambda$ for phase $p$ (see Equation (\ref{eq3})).

In order to reproduce the observed extinction curves with dust models,
the Levenberg-Marquardt method is generally used to perform a grid-search by
minimizing $\chi^{2}$ with the weights
(\hyperlink{WD01}{WD01}; \citealt{Wang}; \hyperlink{A15}{A15}; \citealt{Nozawa}).
\cite{Fritz2011} used $\chi^{2}/{\rm d.o.f}$ to
evaluate the goodness of fitting their observed extinction curve towards
the Galactic center with different dust models.
Therefore, the goodness of our fitting is evaluated by minimizing
\begin{equation}
\chi^{2}/{\rm d.o.f} = \frac{1}{(N_{\rm data} - N_{\rm para})}\sum\limits_{\lambda}
\sum\limits_{p} \frac{[E(\lambda - V)^{\rm mod}_{p} - E(\lambda - V)^{\rm data}_{p}]^2 }{\sigma^2},
\label{eq11}
\end{equation}
%
where at wavelength $\lambda$ for phase $p$, the modeled color excesses
$E(\lambda - V)^{\rm mod}_{p} \equiv A^{\rm mod}_{\lambda,p} - A^{\rm mod}_{{\rm V},p}$\footnote{$A^{\rm mod}_{\lambda,p}$
is calculated by Equation (\ref{eq9}) when
fitting the photometric data, instead, by Equation (\ref{eq8}) when
fitting the spectroscopic data.
};
$E(\lambda - V)^{\rm data}_{p}$ is derived from the observed data by Equation (\ref{eq2})
and (\ref{eq3}), for the photometric data and the spectroscopic data, respectively;
$1/{\sigma^2}$ is the weight, i.e., the uncertainties of data;
$N_{\rm data}$ is the number of observed data points used for fitting,
and $N_{\rm para}$ is the number of adjustable parameters; ${\rm d.o.f} \equiv
N_{\rm data} - N_{\rm para}$ is the degree of freedom.
Since we assume that the silicate and carbonaceous grains have the same dust
size distribution, there are three parameters ($N_{\rm para}=3$) in our dust models:
the size distribution power index $\alpha$,
the exponential cutoff size $a_b$,
and the H column density $\NH$.
The grid of $\alpha$ ranges from 0.4 to 5.0 with a step of 0.1, while
$a_b$ ranges from 0.01 to 0.30 with a step of 0.01.
%
%

\begin{center}
\renewcommand{\arraystretch}{1.2}
\begin{table*}[htbp]
\resizebox{\linewidth}{!}{
\begin{threeparttable}

\caption{\label{tab2} The Results of the Best-fit Dust Models}

\begin{tabular}{cccccccccc}
\toprule
Carbon\tnote{a}  & ${f}_{\rm cs}$\tnote{b} & $\alpha$ & $a_b$ & ${N}_{\rm{H}}$ & ${\chi }^{2}/\mathrm{d.o.f}$\tnote{c} & ${E(B-V)^{\rm mod}}$ & ${A}_{\rm V}$ & ${A}_{\rm Ks}$ & ${R}_{\rm V}$ \\
                 &  &  & ($\mu {\rm{m}}$) & ($\times {10}^{22}\;{\mathrm{cm}}^{-2}$) & & (mag) & (mag) & (mag) & \\
\midrule

{} & 0 & 0.5 & 0.04 & 1.10 & 0.30 & 1.00 & 2.31 & 0.08 & 2.32\\\cline{1-10}
{\multirow{2}{*}{AMC}} & 0.3 & 0.9 & 0.04 & 0.68 & 0.33 & 0.96 & 2.58 & 0.14 & 2.68\\
{} & 0.6 & 1.2 & 0.04 & 0.49 & 0.34 & 0.96 & 2.59 & 0.16 & 2.69\\\cline{1-10}
{\multirow{2}{*}{GRA} } & 0.3 & 0.5 & 0.03 & 0.68 & 0.33 & 0.97 & 2.67 & 0.17 & 2.75\\
{} & 0.6 & 1.0 & 0.03 & 0.47 & 0.33 & 0.96 & 2.73 & 0.19 & 2.84\\

\bottomrule
\end{tabular}
\begin{tablenotes}
\item[a]{Two types of carbonaceous dust are considered in the dust model,
as described in \S\ref{sec:result}. AMC is amorphous carbon, and GRA is graphite.}
\item[b]{$f_{\rm cs}$ is the mass ratio of carbonaceous dust to silicate dust, i.e., $m_{\rm car}/m_{\rm sil}$}.
\item[c]{The values of ${\chi}^{2}/\mathrm{d.o.f}$ are calculated with weights
by considering the uncertainties ($\sigma$).}
\end{tablenotes}
\end{threeparttable}}\\
\end{table*}
\end{center}
%
%

\section{Results}\label{sec:result}
In this work, we calculate the color excesses, $E(X-V)$ of SN 2012cu
in the UV-to-NIR bands (2346$-$21295{\AA}, see \S \ref{sec:obda}) and the
color excess curves, $E(\lambda - V)$ in the optical band (3850$-$9150{\AA}).
We fit the combination of the color excesses and the color excess curves
using the silicate+graphite/amorphous carbon dust models.
The color excesses derived from the photometric data contain 31 points
and cover eight epochs (i.e., $N_{\rm data}=31$ and phase $p=8$).
The color excess curves derived from the spectroscopic data cover 15
epochs ($p=15$, see black lines in Figure \ref{2}).
When fitting with the dust models, each of these 15 color excess
curves is evenly divided into 54 points with a step of 100{\AA},
which yields 810 points ($N_{\rm data}=810$) in total.
Thus, in Equation (\ref{eq11}), we finally take $p=23$ and $N_{\rm data}=841$
for summation and search for the best fitting parameters by minimizing
$\chi^2/{\rm d.o.f}$.

The best-fit results to our dust models are shown in Table \ref{tab2}.
The five rows present the dust models with
different types of carbonaceous component (GRA or AMC)
and mass ratio $f_{\rm cs}$ between carbonaceous and silicate dust.
In Table \ref{tab2}, it is clear that the modeled $\Av$ varies from
2.31 to 2.73$\magni$, $E(B-V)$ ranges from 0.96 to 1.00$\magni$, and $\Rv$
changes from 2.32 to 2.84.
The average values of $\Av$, $E(B-V)$, and $\Rv$ are
approximately $2.6\magni$, $0.97\magni$, and 2.7, respectively,
consistent with those of \hyperlink{A15}{A15} and
\hyperlink{H17}{H17}, i.e, $E(B-V) = 0.98 \sim 1.00 \magni$
and $\Rv = 2.8 \sim 3.0$.

\begin{figure}[htbp]
 \flushleft
 \includegraphics[width=1\linewidth]  {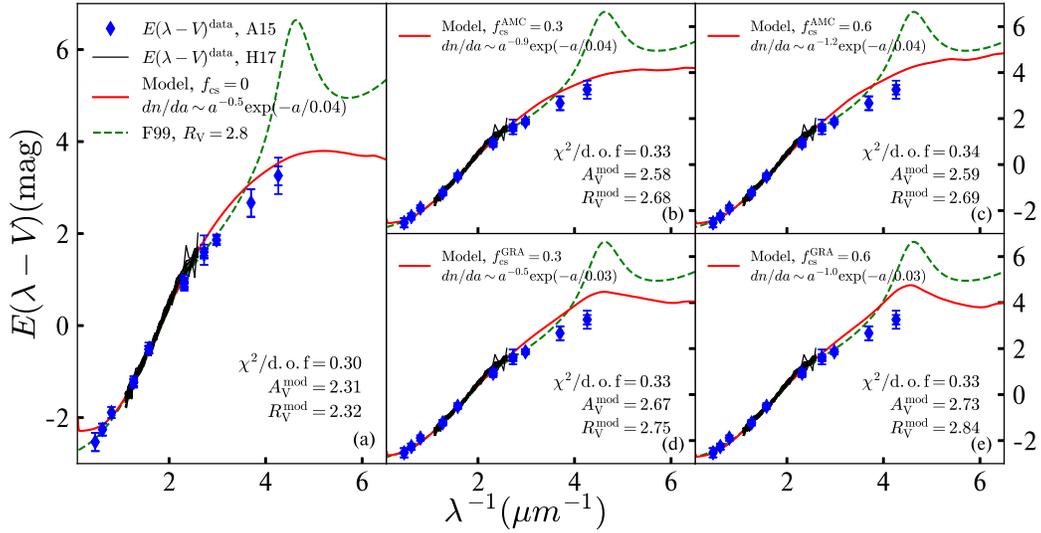}
 \caption{\label{4.sub.1} The modeled $E(\lambda-V)$ of SN 2012cu with $\lambda^{-1}$.
 Blue diamonds present the color excesses directly obtained from photometric
 data, and black solid lines give color excess curves derived
 from spectroscopic data. Red solid lines show the modeled color excess curves
 presented in Table \ref{tab2}. Panel (a) on the left provides the result
 with $f_{\rm cs} = 0$. The panels in the middle (b,d) and the right
 (c,d) columns present the results with $f_{\rm cs} = 0.3$ and 0.6, respectively.
 The results for amorphous carbon (AMC) are shown in the upper middle (b) and the upper right
 (c) panels, and the results for graphite (GRA) are presented in the lower middle (d)
 and the lower right (e) panels. For comparison, in each panel, we overplot the F99
 extinction curves of $\Rv = 2.8$ in green dashed line.}

\end{figure}

Figure \ref{4.sub.1} shows all of the five best-fit results
in terms of $E(\lambda - V)$ with $\lambda^{-1}$.
Blue diamonds show the color excesses derived from the photometric data,
and black solid lines present the color excess curves derived from
all the spectra (see black solid lines in Figure \ref{2}).
Our modeled color excess curves are plotted in red solid lines.
The \hyperlink{F99}{F99} extinction curve of $\Rv = 2.8$ is also shown
by green dashed line for comparison.
In Figure \ref{4.sub.1}, all modeled color excess curves
indicate that the extinction law of SN 2012cu seems much flatter in the
UV bands than the \hyperlink{F99}{F99} law of $\Rv = 2.8$ with
a strong 2175{\AA} bump. By adjusting parameters (e.g., dust
size distribution) of the dust models, the modeled color excess curves
are feasible to display the flatness towards the UV bands even when $\Rv = 2.8$.
Figure \ref{4.sub.1}a presents the reproduced color excess
curve with no carbonaceous dust in the dust model,
which yields the smallest value of $\chi^2/{\rm d.o.f}$.
Because small graphite grains are usually considered
as the candidate carrier of the 2175{\AA} bump (\hyperlink{WD01}{WD01}),
the modeled color excess curves with GRA (Figure \ref{4.sub.1}d
and \ref{4.sub.1}e) show weak 2175{\AA} features and exhibit
conspicuous deviations from the observed color excesses of SN 2012cu
in the UV bands, e.g., in $F225W$ and $F275W$ passbands.
By contrast, the modeled color excess curves with AMC
(Figure \ref{4.sub.1}b and \ref{4.sub.1}c)
coincide with the observed data in these two passbands.

\begin{figure}[htbp]
\flushleft
\includegraphics[width=1\linewidth]  {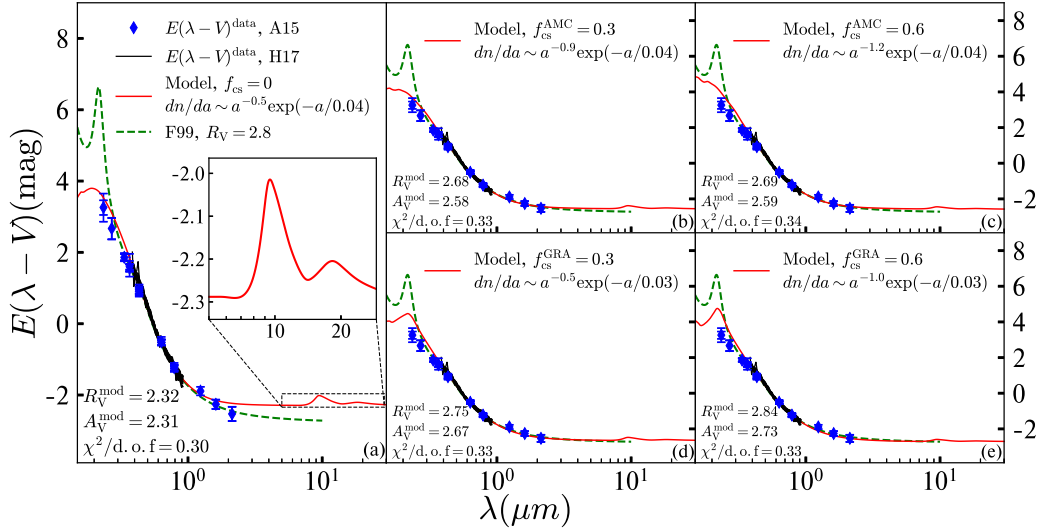}
\caption{\label{4.sub.2}
Same as Figure \ref{4.sub.1}, but for the modeled $E(\lambda-V)$ with
$\lambda$. The inset in panel (a) shows the detailed fitting
curve in the MIR bands.}

\end{figure}

Figure \ref{4.sub.2} shows the modeled color excess curves as
functions of wavelength $\lambda$. These curves extend to the
mid-infrared (MIR) bands and display two absorption
features [see the inset in Figure \ref{4.sub.2}a] around 9.7$\mum$
and 18$\mum$ due to amorphous silicate (\citealt{Li}).
The profiles of these two silicate features
and the corresponding mass ratio of the silicate dust cannot be further
verified due to the lack of observations of SN 2012cu in MIR bandpasses.

\begin{figure}[htbp]
\flushleft
\includegraphics[width=1\linewidth]  {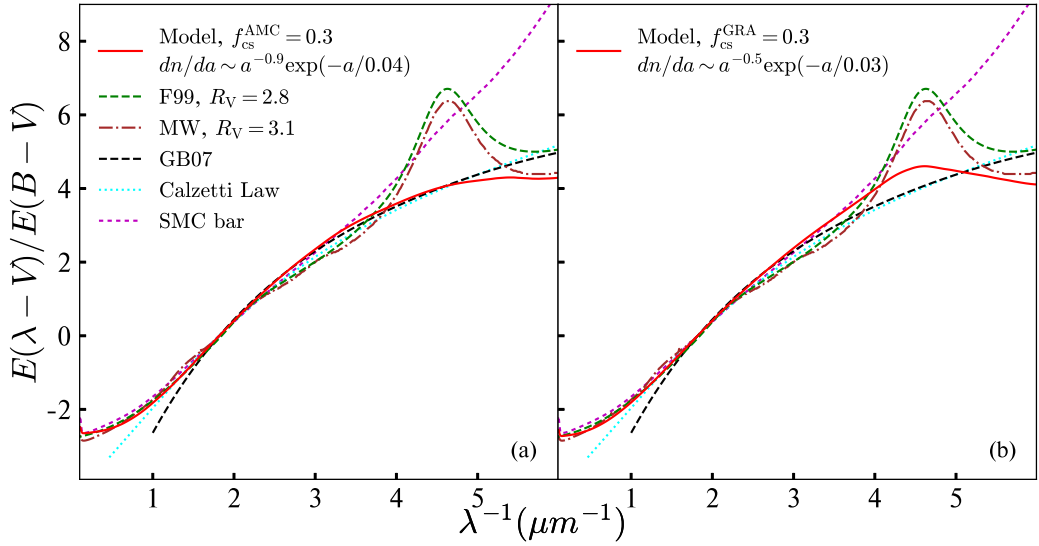}
 \caption{\label{5} Comparison of the normalized color excess curves $E(\lambda-V)/E(B-V)$
 towards SN 2012cu derived from silicate+amorphous carbon dust model
 ($f_{\rm cs}^{\rm AMC} = 0.3$) [Panel (a)] and silicate+graphite dust model
 ($f_{\rm cs}^{\rm GRA} = 0.3$) [Panel (b)].
 Our modeled curves are denoted with red solid lines, while
 the F99 extinction laws of $\Rv = 2.8$ and $\Rv = 3.1$ (MW average) are denoted
 with green dashed lines and brown dot-dashed lines, respectively.
 We also show the extinction curve of the SMC bar (magenta short dashed line),
 the mean reddening curve of AGNs (black dashed line, \citealt{GB07}),
 and the Calzetti attenuation law for starburst galaxies (cyan dotted line, \citealt{Calzetti}).}
\end{figure}

\begin{figure}[htbp]
\vspace{-2cm}
 \center{\includegraphics[width=1\linewidth]  {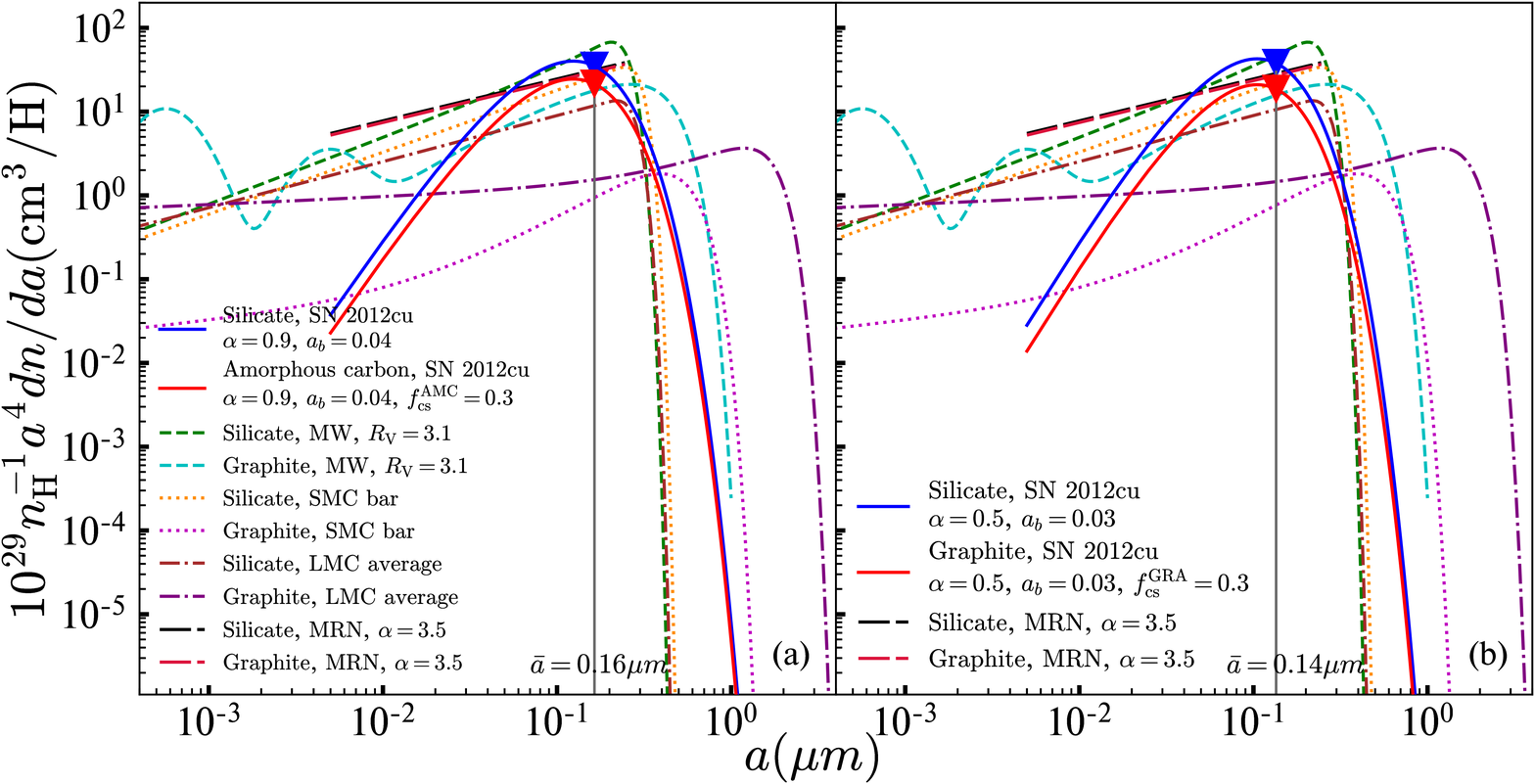}}
 \caption{\label{6} Size distributions of silicate+amorphous carbon dust model (a)
 and silicate+graphite dust model (b) when $f_{\rm cs} = 0.3$. The size distributions of silicate and carbonaceous
 dust are denoted with blue and red solid lines, respectively. The triangles on
 these lines represent the average sizes derived from the curves. For comparison, the WD01
 (\citealt{WD01}) size distributions of silicate and graphite for the average Galactic
 extinction law of $\Rv=3.1$ (green and cyan dashed lines), for the SMC (orange and
 magenta dotted lines), and for the LMC (brown and purple dot-dashed lines) are overplotted,
 respectively. The MRN size distribution with $\alpha = 3.5$ (\citealt{Draine}) is also
 shown by long dashed lines (crimson for silicate and black for graphite).}
\end{figure}

Although our dust model presented by Figure \ref{4.sub.1}a and
\ref{4.sub.2}a provides the minimum value of $\chi^2/{\rm d.o.f}$, it
still produces discernable residuals in the NIR. In addition, since
the detailed abundance of NGC 4772 is not clear, we cannot simply
eliminate the existence of carbonaceous dust in this host galaxy of
SN 2012cu. In NGC 4772, \citet{Ciesla} derived the mass fraction of
polycyclic aromatic hydrocarbons (PAHs) $\approx$ 3.25\%
based on the dust model discussed in \citet{DL07},
even though there is no positive evidence for the presence
of PAHs features. Therefore, the dust models including AMC or GRA grains
should be more reasonable compared to the $f_{\rm cs} = 0$ case.
In Figure \ref{5} and \ref{6},
we present the normalized color excess curves in terms of
$E(\lambda - V)/E(B - V)$ and the corresponding dust size distributions
derived from those models with $f_{\rm cs}^{\rm AMC} = 0.3$ and
$f_{\rm cs}^{\rm GRA} = 0.3$, respectively.

We also adopt the \hyperlink{MRN}{MRN} size distribution to fit
the color excess curves. However, the typical values of
$\chi^{2}/{\rm d.o.f}$ give 0.5 to 0.6, almost twice as the corresponding
values given by our silicate+graphite/amorphous carbon dust model shown in
Table \ref{tab2}. Moreover, the MRN size distribution has even
larger ${\rm d.o.f}$ than that of the power-law function with an
exponential cutoff (\citealt{Kim}), due to the number of adjustable
parameters $N_{\rm para} = 2$, i.e., the power index $\alpha$ and H column
density $\NH$. Therefore, we suggest that the reddening towards
SN 2012cu disfavors the MRN dust size distribution.
%
%
%
%

%
%

\begin{table}
\begin{threeparttable}
\vspace{-1cm}
\begin{center}
\renewcommand{\arraystretch}{0.6}
\vspace{-2cm}
\caption{\label{tab:ext} The Dust Extinction towards SN 2012cu}
\vspace{-0.5cm}
\small
\begin{tabular}{lccccccccccc}
\hline\hline
\multicolumn{2}{c}{}    & \multicolumn{2}{c}{$f_{\rm cs}^{\rm AMC} = 0.3$\tnote{a}}  & \multicolumn{2}{c}{$f_{\rm cs}^{\rm GRA} = 0.3$\tnote{b}}    \\  \cline{3-6}
Band                    & $\lambda$   &$A_{\lambda}$  &   $C_{\rm ext}$/H &$A_{\lambda}$  &   $C_{\rm ext}$/H \\
                        & ($\mum$)    & ($\magni$)    &   (cm$^2/$H)    & ($\magni$)    &   (cm$^2/$H)\\
\hline

Ly edge     &  0.091    &  6.55      &   9.67$\times10^{-22}$&   7.06     &  1.04 $\times10^{-21}$  \\
Ly$\alpha$  &  0.122    &  6.64      &   9.80$\times10^{-22}$&   7.07     & 1.04  $\times10^{-21}$  \\
$UVW2$/UVOT   &  0.203    &  6.61      &   9.76$\times10^{-22}$&   7.04     & 1.03  $\times10^{-21}$  \\
$UVM2$/UVOT   &  0.223    &  6.42      &   9.48$\times10^{-22}$&   7.10     & 1.04  $\times10^{-21}$  \\
$UVW1$/UVOT   &  0.259    &  5.90      &   8.71$\times10^{-22}$&   6.31      & 9.25 $\times10^{-22}$  \\
$F225W$/{\it HST}   &  0.287    &  5.52      &   8.15$\times10^{-22}$&   5.74      & 8.42 $\times10^{-22}$  \\
$F275W$/{\it HST}   &  0.290    &  5.47      &   8.09$\times10^{-22}$&   5.70      & 8.34 $\times10^{-22}$  \\
$F218W$/{\it HST}   &  0.311    &  5.17      &   7.64$\times10^{-22}$&   5.33      & 7.82 $\times10^{-22}$  \\
$F336W$/{\it HST}   &  0.340    &  4.74      &   7.01$\times10^{-22}$&   4.88      & 7.15 $\times10^{-22}$  \\
\emph{u}/SDSS      &  0.355    &  4.54       &   6.71$\times10^{-22}$&   4.66      & 6.83  $\times10^{-22}$  \\
$U$           &  0.365    &  4.41       &  6.51$\times10^{-22}$&   4.52      & 6.63 $\times10^{-22}$  \\
$F438W$/{\it HST}   &  0.433    &  3.61       &  5.33$\times10^{-22}$&   3.71      & 5.44 $\times10^{-22}$  \\
$B$           &  0.440    &  3.54       &  5.23$\times10^{-22}$&   3.64      & 5.34 $\times10^{-22}$  \\
$F467M$/{\it HST}   &  0.468    &  3.26       &  4.82$\times10^{-22}$&   3.36      & 4.93 $\times10^{-22}$  \\
\emph{g}/SDSS      &  0.469    &  3.26       &   4.81$\times10^{-22}$&   3.36      & 4.92  $\times10^{-22}$  \\
$V$           &  0.550    &  2.58       &  3.81$\times10^{-22}$&   2.67      & 3.91 $\times10^{-22}$  \\
$F555W$/{\it HST}   &  0.550    &  2.58       &  3.81$\times10^{-22}$&   2.67      & 3.91 $\times10^{-22}$  \\
\emph{r}/SDSS      &  0.617    &  2.14       &   3.15$\times10^{-22}$&  2.23       & 3.27  $\times10^{-22}$  \\
$F631N$/{\it HST}   &  0.630    &  2.06       &  3.04$\times10^{-22}$&   2.16      & 3.16 $\times10^{-22}$  \\
$R$           &  0.700    &  1.70       &  2.51$\times10^{-22}$&   1.80      & 2.64 $\times10^{-22}$  \\ \
\emph{i}/SDSS      &  0.748    &  1.50       &   2.21$\times10^{-22}$&   1.60      & 2.35  $\times10^{-22}$  \\
$F814W$/{\it HST}   &  0.792    &  1.34       &  1.97$\times10^{-22}$&   1.44      & 2.11 $\times10^{-22}$  \\
$F845M$/{\it HST}   &  0.863    &  1.12       &  1.66$\times10^{-22}$&   1.23      & 1.80 $\times10^{-22}$  \\
\emph{z}/SDSS      &  0.893    &  1.05       &   1.55$\times10^{-22}$&   1.15      & 1.68  $\times10^{-22}$  \\
$I$           &  0.900    &  1.03       &  1.52$\times10^{-22}$&   1.13      & 1.66 $\times10^{-22}$  \\
$J$/2MASS     &  1.235    &  0.51       &  7.51$\times10^{-23}$&   0.58      & 8.57 $\times10^{-23}$  \\
$H$/2MASS     &  1.662    &  0.25       &  3.76$\times10^{-23}$&   0.30      & 4.45 $\times10^{-23}$  \\
$Ks$/2MASS    &  2.159    &  0.14       &  2.10$\times10^{-23}$&   0.17      & 2.52 $\times10^{-23}$  \\
$W1$/WISE     &  3.353    &  0.07       &  1.00$\times10^{-23}$&   0.07      & 1.07 $\times10^{-23}$  \\
$L$           &  3.450    &  0.07       &  9.65$\times10^{-24}$&   0.07      & 1.02 $\times10^{-23}$  \\
$[3.6]$/IRAC&  3.545    &  0.06       &  9.31$\times10^{-24}$&   0.07      & 9.72 $\times10^{-24}$  \\
$[4.5]$/IRAC&  4.442    &  0.05       &  6.80$\times10^{-24}$&   0.05      & 6.70 $\times10^{-24}$  \\
$W2$/WISE     &  4.603    &  0.04       &  6.48$\times10^{-24}$&   0.04      & 6.35 $\times10^{-24}$  \\
$M$           &  4.800    &  0.04       &  6.15$\times10^{-24}$&   0.04      & 5.98 $\times10^{-24}$  \\
$[5.8]$/IRAC&  5.675    &  0.04       &  5.31$\times10^{-24}$&   0.03      & 4.95 $\times10^{-24}$  \\
$[8.0]$/IRAC&  7.760    &  0.05       &  7.31$\times10^{-24}$&   0.04      & 6.43 $\times10^{-24}$  \\
$N$           &  10.600   &  0.15       &  2.21$\times10^{-23}$&   0.15      & 2.15 $\times10^{-23}$  \\
$W3$/WISE     &  11.561   &  0.10       &  1.53$\times10^{-23}$&   0.10      & 1.47 $\times10^{-23}$  \\
$Q$           &  21.000   &  0.06       &  8.55$\times10^{-24}$&   0.06      & 8.72 $\times10^{-24}$  \\
$W4$/WISE     &  22.088   &  0.05       &  7.62$\times10^{-24}$&   0.05      & 7.85 $\times10^{-24}$  \\
\hline
\end{tabular}

\begin{tablenotes}
  \item[a]{The silicate+amorphous carbon dust model with $f_{\rm cs}^{\rm AMC}= 0.3$.}
  \item[b]{The silicate+graphite dust model with $f_{\rm cs}^{\rm GRA} = 0.3$.}

\end{tablenotes}
\end{center}
\end{threeparttable}
\end{table}

%
%
%

%
%

\section{Discussion: The Extinction Law and dust properties towards SN 2012cu}\label{sec:diss}
Since the existence of carbonaceous
dust in NGC 4772 cannot be simply eliminated (see \S \ref{sec:result}),
Figure \ref{5} shows two best-fit results in terms of $E(\lambda - V)/E(B - V)$
when $f_{\rm cs}\neq 0$, in comparison with the extinction curves
calculated with the \hyperlink{F99}{F99} law of $\Rv = 2.8$, the average MW
extinction curve of $\Rv = 3.1$, the extinction curve of the SMC bar
(\hyperlink{WD01}{WD01}), the Calzetti attenuation law (\citealt{Calzetti})
for starburst galaxies, and the mean extinction curve of five mostly reddened
individual AGNs (\citealt{GB07})\footnote{\citet{GB07} derived average extinction
curves of five mostly reddened AGNs in their sample, and they found that
these curves extend flatly towards the far-UV bands without
the 2175{\AA} bump. \citet{Calzetti} derived the attenuation curves of starburst
galaxies, which similarly show the flattening and no appreciable 2175{\AA} bump.
}.
With the silicate+graphite model ($f_{\rm cs}^{\rm GRA} = 0.3$), the modeled extinction
curves exhibit the 2175{\AA} bump produced by small graphite dust and
display discernable deviation from the observed curve in the UV bands.
On the contrary, the silicate+amorphous carbon model ($f_{\rm cs}^{\rm AMC} = 0.3$)
has no 2175{\AA} feature and it well reproduces the extinction curve in
$F225W$ and $F275W$ passbands.

In Figure \ref{5}a, the modeled extinction curves are much flatter in the
UV bands than those of the MW and the SMC bar. It suggests that the size of the dust
in the ISM towards SN 2012cu in NGC 4772 tends to be larger than that of
the MW, and even much larger than that of the SMC bar.
Our modeled extinction curves in the UV bands are nearly similar with the mean
reddening curve of AGNs (\citealt{GB07})\footnote{\citet{Li07} compared the
AGN's extinction curves derived from composite quasar spectra
by \citet{Czerny} and \citet{G04} with the mean extinction curve of AGNs
by \citet{GB07} (see \citealt{Li07} and Figure 3 therein).}.

\hyperlink{A15}{A15} used all measured colors (UV-NIR) to derive the extinction
curves of SN 2012cu with \hyperlink{F99}{F99} extinction law. Their
best-fit result suggests $\Rv = 2.8 \pm 0.1$,
which is shown by green dashed line in Figure \ref{5}.
\hyperlink{H17}{H17} similarly utilized \hyperlink{F99}{F99} law to deredden
SN 2012cu spectra with the best fitting parameters
$E(X - V)=1.0\magni$ and $\Rv=2.95$.
The extinction curves and $\Rv$ of SN 2012cu derived in this work
are, however, based on the detailed dust models for the observed color excesses.
The best-fit values of $\Rv$ for all the cases listed in
Table \ref{tab2} are generally consistent with the studies by
\hyperlink{A15}{A15} and \hyperlink{H17}{H17}.
In Figure \ref{5}b, the presence of a weak 2175{\AA} bump suggests
the existence of graphite as the carbonaceous component merged in the dust model.
Although the \hyperlink{F99}{F99} law of $\Rv = 2.8$ also presents the 2175{\AA}
feature, the profile is too strong to reproduce the flattening of the
UV extinction curve with the same value of $\Rv$.

Figure \ref{6} shows the derived size distributions of silicate
and carbonaceous dust when $f_{\rm cs} = 0.3$,
in comparison with a variety of interstellar dust size distributions.
The average radii of two types of dust grains are presented as $0.16 \mum$ in
Figure \ref{6}a and $0.14 \mum$ in Figure \ref{6}b, respectively.
For the average extinction law of the MW dust with $\Rv = 3.1$,
\citet{Nozawa} presented the average size $\bar{a} \sim 0.01 \mum$
with the dust model consisting of silicate and graphite\footnote{\citet{Nozawa}
calculated the average radii weighted by the size distribution of dust
[see Equation (8) therein], for distinguishing the similarity between an
exponential-like distribution and a lognormal distribution. We adopt the
same equation to derive the average radii for comparing the typical
sizes of dust grains in the MW. Note that the extinction law depends more probably
on the size distribution of dust rather than average radii of dust grains.}.
Compared to \citet{Nozawa},
the average sizes of dust grains derived in our study are larger by approximately a
factor of 10.
According to the modeled extinction laws of SN 2012cu in Figure \ref{5},
we therefore infer that the size distribution of NGC 4772 is biased to large sizes
compared with those of the MW, the LMC and the SMC.

In Table \ref{tab:ext},
we present the modeled extinction correction $A_{\lambda}$ towards SN 2012cu in NGC 4772
with the dust models with $f_{\rm cs}^{\rm AMC}=0.3$
and $f_{\rm cs}^{\rm GRA}=0.3$.
These modeled extinction corrections
cover most of the commonly used astronomical filters.

\section{Conclusion}\label{sec:conc}
In this work, we use the broadband photometric data provided
by \hyperlink{A15}{A15} and optical spectroscopic data obtained from \hyperlink{H17}{H17}
to study the extinction of SN 2012cu.
We adopt silicate+graphite/amorphous carbon dust models to fit the observed color excess
curves of SN 2012cu and derive the extinction as a function of wavelength (i.e., $A_\lambda$).

The best-fit results give the visual extinction towards SN 2012cu of
$\Av \approx 2.6\magni$, the reddening of
$E(B-V)\approx 1.0\magni$, and the total-to-selective extinction ratio
$\Rv$\ $\approx$\ 2.7. The reasonable modeled $A_{\lambda}$ towards SN 2012cu
in NGC 4772 are also presented in Table \ref{tab:ext} for extinction corrections.
The modeled extinction curves towards the SN 2012cu sightline extend
flatly to the far-UV bands with or without a very weak 2175{\AA} feature,
and are much flatter than those of the MW, the LMC, and the SMC.
The flatness of the UV extinction curves suggests a ``grey''
line-of-sight extinction law towards the host AGN galaxy of NGC 4772,
and indicates that the size distribution of dust
in this sightline is skewed to large grains.
The extragalactic extinction
laws are inadequately based on $\Rv$ and probably dependent on
the local interstellar environment.

\section*{Acknowledgements}
We acknowledge Amanullah, R. for providing the SED templates of SN 2011fe.
We thank Prof. Li, A., Dr. Nozawa, T., and the anonymous referees
for their very helpful suggestions/comments.
We are grateful to the helpful discussion during the 10th `Cosmic Dust'
conference held by NAOJ.
This work is supported by NSFC Projects U1631104, and 11533002.



\begin{thebibliography}{}
\bibitem[Amanullah et al.(2014)]{A14} Amanullah, R., et al., 2014,
The Peculiar Extinction Law of SN 2014J Measured with the Hubble Space
Telescope, \href{https://doi.org/10.1088/2041-8205/788/2/L21}{Astrophys. J.}, 788, L21 (6pp).

\bibitem[Amanullah et al.(2015)]{A15} Amanullah, R., et al., 2015,
Diversity in extinction laws of Type Ia supernovae measured between 0.2 and
2 {$\mu$}m, \href{https://doi.org/10.1093/mnras/stv1505}{Mon. Not. R. Astron. Soc.}, 453, 3300-3328. \hypertarget{A15}{(A15)}


\bibitem[Asplund et al.(2009)]{Asplund} Asplund, M., Grevesse, N., Sauval, A.~J.,
Scott, P., 2009, The Chemical Composition of the Sun, \href{https://doi.org/10.1146/annurev.astro.46.060407.145222}{Annu. Rev. Astronom. Astrophys.}, 47, 481-522.

\bibitem[Bohren \& Huffman(1983)]{Bohren} Bohren, C.~F., Huffman, D.~R., 1983, Absorption and scattering of light by small particles,  \href{http://adsabs.harvard.edu/abs/1983asls.book.....B}{New York: Wiley}.

\bibitem[Brown et al.(2015)]{Brown} Brown, P.~J., et al., 2015, Swift Ultraviolet Observations of Supernova 2014J in M82: Large Extinction from Interstellar Dust, \href{https://doi.org/10.1088/0004-637X/805/1/74}{Astrophys. J.}, 805, 74 (13pp).

\bibitem[Buton et al.(2013)]{Buton} Buton, C., et al., 2013,
Atmospheric extinction properties above Mauna Kea from the Nearby SuperNova
Factory spectro-photometric data set, \href{https://doi.org/10.1051/0004-6361/201219834}{Astronom. Astrophys.}, 549, A8 (21pp).

\bibitem[Calzetti et al.(1994)]{Calzetti} Calzetti, D., Kinney, A.~L.,
Storchi-Bergmann, T., 1994, Dust extinction of the stellar continua in
starburst galaxies: The ultraviolet and optical extinction law, \href{https://doi.org/10.1086/174346}{Astrophys. J.}, 429,
582-601.


\bibitem[Cardelli et al.(1989)]{Cardelli} Cardelli, J.~A., Clayton, G.~C., Mathis,
J.~S., 1989, The relationship between infrared, optical, and ultraviolet
extinction, \href{https://doi.org/10.1086/167900}{Astrophys. J.}, 345, 245-256. \hypertarget{CCM89}{(CCM89)}

\bibitem[Ciesla et
al.(2014)]{Ciesla} Ciesla, L., et al., 2014, Dust spectral energy distributions of nearby galaxies: an insight from the Herschel Reference Survey, \href{https://doi.org/10.1051/0004-6361/201323248}{Astronom. Astrophys.}, 565, A128 (33pp).

\bibitem[Cikota et al.(2016)]{Cikota} Cikota, A., Deustua, S., Marleau, F., 2016, Determining Type Ia Supernova Host Galaxy Extinction Probabilities and a Statistical Approach to Estimating the Absorption-to-reddening Ratio R$_{V}$,  \href{https://doi.org/10.3847/0004-637X/819/2/152}{Astrophys. J.}, 819, 152 (13pp).

\bibitem[Clayton et al.(2003)]{Clayton} Clayton, G.~C., et al., 2003,
Dust Grain Size Distributions from MRN to MEM, \href{https://doi.org/10.1086/374316}{Astrophys. J.}, 588, 871-880.

\bibitem[Czerny et al.(2004)]{Czerny} Czerny, B., Li, J., Loska, Z., Szczerba, R., 2004, Extinction due to amorphous carbon grains in red quasars from the Sloan Digital Sky Survey, \href{https://doi.org/10.1111/j.1365-2966.2004.07590.x}{Mon. Not. R. Astron. Soc.}, 348, L54-L57.

\bibitem[Djupvik \& Andersen(2010)]{Djupvik} Djupvik, A.~A., Andersen, J., 2010, The Nordic Optical Telescope,  \href{https://doi.org/10.1007/978-3-642-11250-8_21}{Astrophys. Space Sci. Proc.}, 14, 211.

\bibitem[Draine(2011)]{Draine2011} Draine, B.~T., 2011, Physics
of the Interstellar and Intergalactic Medium,
\href{http://adsabs.harvard.edu/abs/2011piim.book.....D}{Princeton
University Press}, ISBN: 978-0-691-12214-4.

\bibitem[Draine
\& Lee(1984)]{Draine} Draine, B.~T., Lee, H.~M., 1984,
Optical properties of interstellar graphite and silicate grains, \href{https://doi.org/10.1086/162480}{Astrophys. J.}, 285, 89-108.

\bibitem[Draine
\& Li(2007)]{DL07} Draine, B.~T., Li, A., 2007, Infrared Emission from Interstellar Dust. IV. The Silicate-Graphite-PAH Model in the Post-Spitzer Era, \href{https://doi.org/10.1086/511055}{Astrophys. J.}, 657, 810-837.

\bibitem[Fitzpatrick(1999)]{F99} Fitzpatrick, E.~L., 1999, Correcting for the
Effects of Interstellar Extinction, \href{https://doi.org/10.1086/316293}{Publ. Astronom. Soc. Paci.}, 111, 63-75. \hypertarget{F99}{(F99)}

\bibitem[Fitzpatrick \& Massa(2007)]{FM07} Fitzpatrick, E.~L., \& Massa, D., 2007, An Analysis of the Shapes of Interstellar Extinction Curves. V. The IR-through-UV Curve Morphology, \href{https://doi.org/10.1086/518158}{Astrophys. J.}, 663, 320-341.

\bibitem[Folatelli et al.(2010)]{Folatelli} Folatelli, G., et al., 2010, The Carnegie Supernova
Project: Analysis of the First Sample of Low-Redshift Type-Ia Supernovae,
\href{https://doi.org/10.1088/0004-6256/139/1/120}{Astronom. J.}, 139, 120-144.

\bibitem[Foley et al.(2014)]{Foley} Foley, R.~J., et al., 2014,
Extensive HST ultraviolet spectra and multiwavelength observations of SN
2014J in M82 indicate reddening and circumstellar scattering by typical
dust, \href{https://doi.org/10.1093/mnras/stu1378}{Mon. Not. R. Astron. Soc.}, 443, 2887-2906.

\bibitem[Fritz et al.(2011)]{Fritz2011} Fritz, T.~K., et al., 2011, Line Derived Infrared Extinction toward
the Galactic Center, \href{https://doi.org/10.1088/0004-637X/737/2/73}{Astrophys. J.}, 737, 73 (21pp).

\bibitem[Gao et al.(2015)]{Gao} Gao, J., et al., 2015, Physical Dust
Models for the Extinction toward Supernova 2014J in M82, \href{https://doi.org/10.1088/2041-8205/807/2/L26}{Astrophys. J.}, 807, L26 (6pp).

\bibitem[Gao et al.(2013)]{Gao13} Gao, J., Li, A., Jiang, B.~W., 2013, Modeling the infrared extinction toward the galactic center, \href{http://adsabs.harvard.edu/abs/2013EP\%26S...65.1127G}{Earth Planets Space}, 65, 1127-1132.


\bibitem[Gaskell
\& Benker(2007)]{GB07} Gaskell, C.~M., Benker, A.~J., 2007, AGN Reddening and Ultraviolet Extinction Curves from Hubble Space Telescope Spectra, \href{http://adsabs.harvard.edu/abs/2007arXiv0711.1013G}{arXiv:0711.1013}.

\bibitem[Gaskell et al.(2004)]{G04} Gaskell, C.~M., Goosmann, R.~W., Antonucci, R.~R.~J., Whysong, D.~H., 2004, The Nuclear Reddening Curve for Active Galactic Nuclei and the Shape of the Infrared to X-Ray Spectral Energy Distribution, \href{https://doi.org/10.1086/423885}{Astrophys. J.}, 616, 147-156.

\bibitem[Goobar(2008)]{Goobar08} Goobar, A., 2008, Low R$_{V}$
from Circumstellar Dust around Supernovae,
\href{https://doi.org/10.1086/593060}{Astrophys. J.}, 686, L103-L106.

\bibitem[Goobar et al.(2014)]{Goobar} Goobar, A., et al., 2014, The Rise of SN 2014J in the Nearby Galaxy M82, \href{https://doi.org/10.1088/2041-8205/784/1/L12}{Astrophys. J.}, 784, L12 (6pp).

\bibitem[Gordon et al.(2003)]{Gordon} Gordon, K.~D., et al., 2003, A
Quantitative Comparison of the Small Magellanic Cloud, Large Magellanic
Cloud, and Milky Way Ultraviolet to Near-Infrared Extinction Curves, \href{https://doi.org/10.1086/376774}{Astrophys. J.}, 594, 279-293.

\bibitem[Hamuy et al.(1993)]{Hamuy} Hamuy, M., Phillips,
M.~M., Wells, L.~A.,
 Maza, J., 1993, K Corrections for type IA supernovae, \href{http://adsabs.harvard.edu/abs/1993PASP..105..787H}{Publ. Astronom. Soc. Paci.}, 105, 787-793.

\bibitem[Haynes et al.(2000)]{Haynes} Haynes, M.~P., et al., 2000,
Kinematic Evidence of Minor Mergers in Normal SA Galaxies: NGC 3626, NGC
3900, NGC 4772, and NGC 5854, \href{https://doi.org/10.1086/301457}{Astronom. J.}, 120, 703-727.


\bibitem[Ho et al.(1997)]{Ho} Ho, L.~C., Filippenko, A.~V., Sargent, W.~L.~W.,
Peng, C.~Y., 1997, A Search for ``Dwarf'' Seyfert Nuclei. IV. Nuclei with
Broad H{$\alpha$} Emission, \href{https://doi.org/10.1086/313042}{Astrophys. J. Suppl.}, 112, 391-414.


\bibitem[Huang et al.(2017)]{H17} Huang, X., et al., 2017, The
Extinction Properties of and Distance to the Highly Reddened Type IA
Supernova 2012CU, \href{https://doi.org/10.3847/1538-4357/836/2/157}{Astrophys. J.}, 836, 157 (18pp). \hypertarget{H17}{(H17)}


\bibitem[Itagaki et al.(2012)]{Itagaki} Itagaki, K., et al., 2012,
Supernova 2012cu in NGC 4772 = Psn J12532935+0209390, \href{http://adsabs.harvard.edu/abs/2012CBET.3146....1I}{Central Bureau
Electronic Telegrams}, 3146, 1.

\bibitem[Kim et al.(1994)]{Kim} Kim, S.-H., Martin, P.~G., Hendry, P.~D., 1994, The size distribution of interstellar dust particles as determined from extinction, \href{https://doi.org/10.1086/173714}{Astrophys. J.}, 422, 164-175.

\bibitem[Li(2007)]{Li07} Li, A., 2007, Dust in Active
Galactic Nuclei, \href{http://adsabs.harvard.edu/abs/2007ASPC..373..561L}{ASP Conf. Ser.}, 373, 561--572.

\bibitem[Li
\& Draine(2001)]{Li} Li, A., Draine, B.~T., 2001, Infrared Emission from Interstellar Dust. II. The Diffuse Interstellar Medium, \href{https://doi.org/10.1086/323147}{Astrophys. J.}, 554, 778-802.


\bibitem[Marion et al.(2012)]{Marion} Marion, G.~H., Milisavljevic, D., Rines, K.,
Wilhelmy, S., 2012, Supernova 2012cu in NGC 4772 = PSN J12532935+0209390.,
\href{http://adsabs.harvard.edu/abs/2012CBET.3146....2M}{Central Bureau Electronic Telegrams}, 3146, 2.


\bibitem[Mathis et al.(1977)]{MRN} Mathis, J.~S., Rumpl, W., Nordsieck, K.~H.,
1977, The size distribution of interstellar grains, \href{https://doi.org/10.1086/155591}{Astrophys. J.}, 217, 425-433. \hypertarget{MRN}{(MRN)}

\bibitem[Mazzali et al.(2014)]{Mazzali} Mazzali, P.~A., et al., 2014,
Hubble Space Telescope spectra of the Type Ia supernova SN 2011fe: a tail
of low-density, high-velocity material with Z $<$ Z$_{\odot}$, \href{https://doi.org/10.1093/mnras/stu077}{Mon. Not. R. Astron. Soc.}, 439, 1959-1979.


\bibitem[Nobili
\& Goobar(2008)]{Nobili} Nobili, S., Goobar, A., 2008, The colour-lightcurve shape relation of type Ia supernovae and the reddening law, \href{https://doi.org/10.1051/0004-6361:20079292}{Astronom. Astrophys.}, 487, 19-31.

\bibitem[Nozawa(2016)]{Nozawa} Nozawa, T., 2016, Properties of interstellar dust
responsible for extinction laws with unusually low total-to-selective
extinction ratios of R$_{V}$=1-2, \href{https://doi.org/10.1016/j.pss.2016.08.006}{Planet. Space Sci.}, 133, 36-46.

\bibitem[Nugent et al.(2011)]{Nugent} Nugent, P.~E., et al., 2011,
Supernova SN 2011fe from an exploding carbon-oxygen white dwarf star,
\href{https://doi.org/10.1038/nature10644}{Nat.}, 480, 344-347.

\bibitem[Patat et al.(2007)]{Patat07} Patat, F., et al., 2007, Detection of Circumstellar Material in a
Normal Type Ia Supernova,
\href{https://doi.org/10.1126/science.1143005}{Sci.}, 317, 924-926.

\bibitem[Patat et al.(2013)]{Patat} Patat, F., et al., 2013,
Multi-epoch high-resolution spectroscopy of SN 2011fe. Linking the
progenitor to its environment, \href{https://doi.org/10.1051/0004-6361/201118556}{Astronom. Astrophys.}, 549, A62 (10pp).

\bibitem[Pereira et al.(2013)]{Pereira} Pereira, R., et al.,  2013,  Spectrophotometric time series of SN 2011fe from the Nearby Supernova Factory, \href{https://doi.org/10.1051/0004-6361/201221008}{Astronom. Astrophys.}, 554, A27 (22pp).

\bibitem[Perlmutter et al.(1999)]{Perlmutter} Perlmutter, S., et al., 1999, Measurements of {$\Omega$} and {$\Lambda$} from 42 High-Redshift Supernovae, \href{https://doi.org/10.1086/307221}{Astrophys. J.}, 517, 565-586.

\bibitem[Riess et al.(1998)]{Riess98} Riess, A.~G., et al., 1998, Observational Evidence from Supernovae
for an Accelerating Universe and a Cosmological Constant,
\href{https://doi.org/10.1086/300499}{Astronom. J.}, 116, 1009-1038.

\bibitem[Riess et al.(2016)]{Riess16} Riess, A.~G., et al., 2016, A 2.4\% Determination of the Local Value of the Hubble Constant, \href{https://doi.org/10.3847/0004-637X/826/1/56}{Astrophys. J.}, 826, 56 (31pp).

\bibitem[Rouleau
\& Martin(1991)]{Rouleau} Rouleau, F., Martin, P.~G., 1991,
Shape and clustering effects on the optical properties of amorphous carbon, \href{https://doi.org/10.1086/170382}{Astrophys. J.}, 377, 526-540.

\bibitem[Schlafly
\& Finkbeiner(2011)]{Schlafly} Schlafly, E.~F., Finkbeiner, D.~P., 2011,
Measuring Reddening with Sloan Digital Sky Survey Stellar Spectra and Recalibrating SFD, \href{https://doi.org/10.1088/0004-637X/737/2/103}{Astrophys. J.}, 737, 103 (13pp).

\bibitem[Stritzinger et al.(2002)]{Stritzinger} Stritzinger, et al., 2002, Optical Photometry of the Type
Ia Supernova 1999ee and the Type Ib/c Supernova 1999ex in IC 5179,
\href{http://adsabs.harvard.edu/abs/2002AJ....124.2100S}{Astronom. J.}, 124,
2100-2117.

\bibitem[Tully et al.(2008)]{Tully08} Tully, R.~B., et al., 2008, Our Peculiar Motion Away from
the Local Void, \href{https://doi.org/10.1086/527428}{Astrophys. J.}, 676,
184-205.

\bibitem[Tully et al.(2009)]{Tully} Tully, R.~B., et al., 2009, The Extragalactic Distance Database,
\href{http://adsabs.harvard.edu/abs/2009AJ....138..323T}{Astronom. J.}, 138,
323-331.

\bibitem[Wang(2005)]{Wang05} Wang, L., 2005, Dust around
Type Ia Supernovae, \href{https://doi.org/10.1086/499053}{Astrophys. J.}, 635,
L33-L36.

\bibitem[Wang et
al.(2014)]{Wang} Wang, S., Li, A., Jiang, B.~W., 2014, Modeling the infrared interstellar extinction,  \href{https://doi.org/10.1016/j.pss.2014.03.018}{Planet. Space Sci.}, 100, 32-39.

\bibitem[Weingartner
\& Draine(2001)]{WD01} Weingartner, J.~C., Draine, B.~T., 2001,
Dust Grain-Size Distributions and Extinction in the Milky Way,
Large Magellanic Cloud, and Small Magellanic Cloud, \href{https://doi.org/10.1086/318651}{Astrophys. J.}, 548, 296-309. \hypertarget{WD01}{(WD01)}

\bibitem[Yang et al.(2017)]{Yang} Yang, Y., et al., 2017, Interstellar-medium Mapping in M82 through Light Echoes around Supernova 2014J, \href{https://doi.org/10.3847/1538-4357/834/1/60}{Astrophys. J.}, 834, 60 (15pp).

\bibitem[Zhang et al.(2016)]{Zhang} Zhang, K., et al., 2016, Optical Observations of the Type Ia Supernova SN2011fe in M101 for Nearly 500 Days, \href{https://doi.org/10.3847/0004-637X/820/1/67}{Astrophys. J.}, 820, 67 (17pp).

\end{thebibliography}
\end{document}